\documentclass[acmlarge]{acmart}

\makeatletter
\newcommand{\myconfshort}{\acmConference@shortname}
\newcommand{\myconffull}{\acmConference@name}
\newcommand{\myconfdate}{\acmConference@date}
\newcommand{\myconfloc}{\acmConference@venue}
\AtBeginDocument{
  \fancypagestyle{firstpagestyle}{
    \fancyhead{}%
    \fancyfoot[C]{}%
  }
  \fancyhf{}
  \fancyhead[LO]{\@headfootfont\shorttitle}%
  \fancyhead[RE]{\@headfootfont\@shortauthors}%
  \fancyhead[LE]{\@headfootfont\footnotesize \myconfshort, \myconfdate, \myconfloc}%
  \fancyhead[RO]{\@headfootfont\footnotesize \myconfshort, \myconfdate, \myconfloc}%
  \fancyfoot[C]{}%
}
\makeatother
\acmBooktitle{\conffull\@ (\confshort), \confdate, \confloc}

\usepackage{enumitem}

\usepackage{tikz}
\usetikzlibrary{shapes, backgrounds}

\AtBeginDocument{%
  }

\copyrightyear{2026}
\acmYear{2026}
\setcopyright{cc}
\setcctype{by}
\acmConference[FAccT '26]{The 2026 ACM Conference on Fairness, Accountability, and Transparency}{June 25--28, 2026}{Montreal, QC, Canada}
\acmBooktitle{The 2026 ACM Conference on Fairness, Accountability, and Transparency (FAccT '26), June 25--28, 2026, Montreal, QC, Canada}
\acmDOI{10.1145/3805689.3812396}
\acmISBN{979-8-4007-2596-8/2026/06}

\begin{document}
\newcommand{\todo}[1]{\textcolor{red}{(to-do) #1}}
\newcommand\rev[1]{\textcolor{blue}{#1}}

\title{The Quiet Path from Seemingly Minor Design Errors to Workplace AI Incidents}


\author{Julia De Miguel Velázquez}
\orcid{0009-0000-5724-8275}
\affiliation{%
  \institution{King's College London}
  \city{London}
  \country{United Kingdom}
}
\email{julia.de_miguel_velazquez@kcl.ac.uk}

\author{Sanja Šćepanović}
\orcid{0000-0002-1534-8128}
\affiliation{
\institution{Nokia Bell Labs}
\city{Cambridge}
\country{United Kingdom}}
\affiliation{\institution{University of Oxford}
\city{Oxford}
\country{United Kingdom}}
\email{sanja.scepanovic@nokia-bell-labs.com}

\author{Andrés Gvirtz}
\orcid{0000-0001-5815-7164}
\affiliation{%
  \institution{King's College London}
  \city{London}
  \country{United Kingdom}
}
\email{andres.gvirtz@kcl.ac.uk}

\author{Daniele Quercia}
\orcid{0000-0001-9461-5804}
\affiliation{\institution{Nokia Bell Labs} 
\city{Cambridge}
\country{United Kingdom}}
\affiliation{\institution{Politecnico di Torino}
\city{Turin}
\country{Italy}}
\email{quercia@cantab.net}

\renewcommand{\shortauthors}{De Miguel Velázquez et al.}

\begin{abstract}
Recent human-computer interaction (HCI) research has revealed a widespread misalignment between how developers design workplace artificial intelligence (AI) systems, and what workers actually need from them. Yet, little research has examined the effects of this gap, or how it may cause harm. 
We analyzed 1,524 reports of incidents in which AI systems were used to perform 171 occupational tasks across 12 industry sectors. Using an Large Language Model (LLM)-as-an-expert approach, we extracted the main traits of the AI systems involved in those incidents using an established framework of twelve traits. We then compared them with the traits that 202 workers highly familiar with those tasks would have preferred.
We found that as many as 83\% of workplace incidents stem from worker-AI misalignments. In most cases, workers wanted systems that are precise, insightful, or personal, but instead received systems that are basic, simple, or general. 
Over the years, fast AI caused a considerable number of incidents, yet these declined, and imaginative AI, with the mass introduction of generative AI, started to cause incidents. We also compared the traits causing the incidents with the traits that 197 developers building AI systems for those tasks would have preferred. If the traits causing the incidents were the same as those designed by developers, then developers may be responsible for those incidents. We found that 74\% of task misalignments could be attributed to developers who tended to overfocus on efficiency and speed, especially for systems performing tasks in people-facing occupations such as those in the human resources sector. Our results call for design interventions that better align AI development with workers' needs, as without such corrections, workplace AI incidents are likely to persist, causing the invisible erosion of worker agency and organizational productivity.

\end{abstract}




\received{13 January 2026}
\received[revised]{25 February 2026}
\received[accepted]{15 April 2026}


\maketitle

\section{Introduction}


AI companies framed AI for workers and organisations as a tool that would reduce repetitive work, and increase productivity~\cite{bareis2022talking, brynjolfsson2025generative}. Now, workers use these systems for their everyday tasks, from software development to legal research~\cite{brachman2025current,handa2025economic,shao_future_2025}. Yet, in practice, AI adoption is not smooth, nor does it function as intended; workers often compensate by doing extra labor~\cite{fox_patchwork_2023,nedzhvetskaya_no_2024,lee2025impact}. Work in human-computer interaction (HCI), and science and technology studies (STS) may explain why: developers design systems with flawed assumptions about worker needs~\cite{ranjit_are_we,raji2020closing,suchman_making_1995,forsythe1993engineering}.  
To address this, systems should align with actual worker needs and practices to meaningfully support their work tasks~\cite{workshop_work_chi,awumey2024systematic,acemoglu2023can}. 
For example, a lawyer drafting contracts needs precision, 
while a worker brainstorming a topic needs creative and exploratory output, even if inaccurate.
An AI system is \emph{misaligned} when it performs a task with traits that differ from the traits that workers would have preferred for the system.



The problem is that seemingly minor misalignments may lead to incidents, from 
biased hiring~\cite{wang_we_2023} 
to stress from algorithmic management systems \cite{lynn_regulating_2025, nedzhvetskaya_no_2024}. 
Yet, these incidents have not informed design processes so far, representing a critical but untapped source of evidence. We argue that systematically documenting workplace AI incidents offers a unique opportunity to extract design lessons that could better align AI systems with workers' needs. In tackling this opportunity, we make two main contributions:

\begin{enumerate}[leftmargin=*]
\item 
\textbf{We developed an evaluation framework of LLM rubrics and surveys to analyze workplace AI incidents caused by misaligned AI systems~(\S\ref{sec:methods}).}  
We analysed 1,256 incidents from the AI Incident Database (AIID) \cite{mcgregor_preventing_2021} from 2013-2025.
We used a validated LLM-approach to identify workplace incidents in which AI systems were used to perform 171 occupational tasks across 12 industry sectors (e.g., drafting legal documents). 
Drawing on 12 widely-used pairs of opposing psychological AI traits~\cite{fears_ai_dong_2024}, we gathered 202 worker and 197 developer preferences about how their work tasks should ideally be exposed to AI, building on previous work~\cite{ranjit_are_we,shao_future_2025,dong_fears_2024}.
Our framework was able to determine whether an AI trait misalignment caused the incident (e.g., the AI was too \emph{imaginative}, but the worker wanted it to be \emph{practical}),
and whether the incident may be attributed to developers (e.g., the developer designed it to be \emph{imaginative}). 


\item 
\textbf{We quantify the extent to which such incidents are caused by misaligned AI systems (\S\ref{sec:results})}.  We found AI trait misalignment with workers’ needs plays a major role in workplace incidents, causing 83.4\% of them. 
Misalignment often happens when workers want \emph{precise, insightful}, or \emph{personal} systems but receive \emph{basic, simple}, or \emph{general} ones. 
We further found these results vary by sector, e.g., workers in the legal sector are involved in incidents with \emph{imaginative} AI, while human resources with \emph{fast} AI.
However, over time, incidents involving \emph{fast} AI declined, while those involving \emph{imaginative} AI increased, likely due to the rise of generative AI.
We found that most misaligned tasks could be attributed to the developers’ design decisions (74\%).
Developers differ most from workers because they prioritize traits related to efficiency and speed, i.e., \emph{basic} \emph{vs.} \emph{precise}, and \emph{fast} \emph{vs.} \emph{explainable}.
\end{enumerate}


The implications of these results are theoretical and practical  (\S\ref{sec:discussion}). Theoretically, we conceptualise AI incident through the lens of trait misalignment, a shift that will allow researchers to understand failures from a HCI angle. Practically, trait misalignment is a risk factor and should be included in risk assessments. Structural interventions are needed at the design stage to address the causes of developer misalignment, from micro (developers' technical training and understanding of experiences of people from different backgrounds, who may be affected by AI in different ways), to meso (organizational pressures around productivity and automation), to macro (geopolitical and economic forces that deprioritize worker needs). 
To support researchers in advancing this research direction, we have publicly released our approach at \textbf{\url{https://social-dynamics.net/ai-impact/incidents/}}.

\section{Related Work}\label{sec:rw}

Our work builds on a rich discussion in the HCI and AI ethics communities (i.e., FAccT, CSCW, CHI) on AI misalignment with worker needs (\S\ref{sec:rw_alignment}), and AI incidents in the workplace (\S\ref{sec:rw_incidents}), taking a worker-centric approach. 

\subsection{AI Misalignment with Worker Needs}\label{sec:rw_alignment}

HCI is concerned with closing the gap between user mental models and system optimisation goals, to avoid negative consequences~\cite{norman2013design,weisz_design_2024}.
For example, users may expect human-like reasoning from LLMs, though these systems reflect statistical patterns~\cite{bender_dangers_2021}. 
AI alignment aims to make systems `behave' in line with user intentions, preferences, and values~\cite{gabriel_ethics_2024, ji_ai_2025, kirk2025human}. 
These intentions can be operationalised as the psychological traits users prefer the systems to have \cite{fears_ai_dong_2024}. 
Early frameworks emphasised helpfulness, honesty, and harmlessness ~\cite{askell2021general}.
However, critical approaches suggests that alignment is context-dependent \cite{gabriel_ethics_2024}. 
For instance, an honest AI system may be appropriate in managerial contexts, but not in sensitive healthcare conversations where care is needed~\cite{fears_ai_dong_2024}. 

Research on workplace AI has found that technology adoption (i.e., tasks exposed to AI systems) depends on (mis)alignment between workers, systems, and tasks \cite{ammenwerth_it-adoption_2006}.
Recent work found that LLMs meant to summarise reports in clinical settings failed due to a misalignment between the system being too \textit{structured} while the clinicians wanted it to be \textit{flexible} \cite{kupferschmidt_write_2025}.
\citet{fox_patchwork_2023} conceptualised `patchwork', referring to extra human labour to account for what AI claimed to do and what it actually accomplished. 
Altogether, these studies suggest that, in the workplace, we should evaluate AI's efficacy (`does it even work?'), not just its efficiency~\cite{fox_patchwork_2023,kupferschmidt_write_2025}. 


To tackle this, HCI work turns to the design-stage, and how developers address worker needs~\cite{suchman_making_1995}.
Some studies claim worker experience in the design of systems remains undervalued, e.g., in feminised jobs~\cite{kawakami2026ai}.
\citet{ranjit_are_we} compared worker and developer preferences  about how their job tasks should be exposed to AI.
They found systematic AI trait misalignment: developers emphasized politeness, strictness, and imagination in system design, while workers preferred systems that are straightforward, tolerant, and practical.
This showed the importance of developer and worker collaboration to ensure AI systems align with worker needs~\cite{sadeghian2025workai, suchman_making_1995}.

\subsection{Analysis of AI Incidents in the Workplace}\label{sec:rw_incidents} 

Workplace AI has been linked to incidents across sectors. For instance, hiring managers have used AI resume screeners, producing biased outcomes that disadvantage job-seekers~\cite{wang_we_2023,ingber_emotion_2025}. 
Welfare caseworkers have used eligibility AI tools that misclassify vulnerable populations, denying benefits or delaying critical support; sometimes these systems have been turned down, as in the Netherlands~\cite{scott_algorithmic_2022}. 
Algorithmic management tools in gig and retail work have also imposed strict schedules, and intensified labor, generating stress and reducing meaningful engagement with tasks~\cite{lynn_regulating_2025,nedzhvetskaya_no_2024,awumey_systematic_2024,gausen_framework_2024}. 
Some harms may occur regardless of workers’ intentions~\cite{wang_we_2023}.

Systematically documenting and analysing real-world failures is essential for building safer systems in high-stakes domains, such as aviation and cybersecurity~\cite{turri_why_2023}. 
AI safety research responded with the creation of incident databases, including the AIID~\cite{mcgregor_preventing_2021}, the OECD’s AI Incident Monitor ~\cite{oecd_towards_2025}, and the AIAAIC~\cite{aiaaic_repository}. 
These databases have been key for exploring AI incidents~\cite{demiguel_incidents_2024, incidentsmobile,lee2024deepfakes}, and raising awareness of AI risks across developers~\cite{feffer_ai_2023}, and the public~\cite{bogucka_atlas_2024}. 
Recent work suggests that, while AI incident analyses are rich, they tend to prioritise documenting harm outcomes over examining the upstream design decisions that shaped system behaviour~\cite{turri_why_2023}. 

Instead, sociotechnical approaches to incident analysis emphasise understanding organisational and design decisions that create harm-prone conditions, rather than blaming individual workers, to inform safety guidelines~\cite{leveson_engineering_2011,elish_moral_2019,selbst_fairness_2019}. 
This approach highlights three factors: miscalibration, when design decisions fail to communicate system capabilities~\cite{okamura_adaptive_2020, lee_trust_2004, jacobs_how_2021}; automation surprise, when systems behave differently than workers expect~\cite{woods_learning_2000, dekker_behind_2006, sarter_automation_1997}; 
and systems thinking, which explains how localised misalignment can cascade into significant incidents in complex, tightly coupled systems ~\cite{perrow_normal_1984,rasmussen_risk_1997}, like AI~\cite{bianchi2023artificial}.
From this view, AI trait misalignment may constitute a condition that triggers such failures.

\vspace{0.5em}
\noindent
\textbf{Research gap.} 
Design choices are misaligned with worker needs, and incidents might be rooted in design choices. 
Yet, these literatures remain disconnected: prior work has overlooked whether misalignment between design and worker needs translates into workplace incidents. 
We address this gap by analyzing workplace AI incidents through the lens of AI trait misalignment, and comparing them to worker needs and developer design choices. 



\section{Research Design}\label{sec:methods}
Our work asks two research questions (RQ): 

\begin{enumerate}[label=\textbf{RQ\arabic*.}, leftmargin=*, labelindent=0pt]
    \item 
    To what extent are workplace AI incidents caused by misaligned AI?
    \item 
    When trait misalignment results in incidents, how often could it be attributed to developers \emph{vs.} other causes?
\end{enumerate}

\begin{figure}[!t]
    \centering
    \includegraphics[
        width=1\columnwidth,
        trim=3.4cm 7.37cm 3.8cm 8.50cm,
        clip
    ]{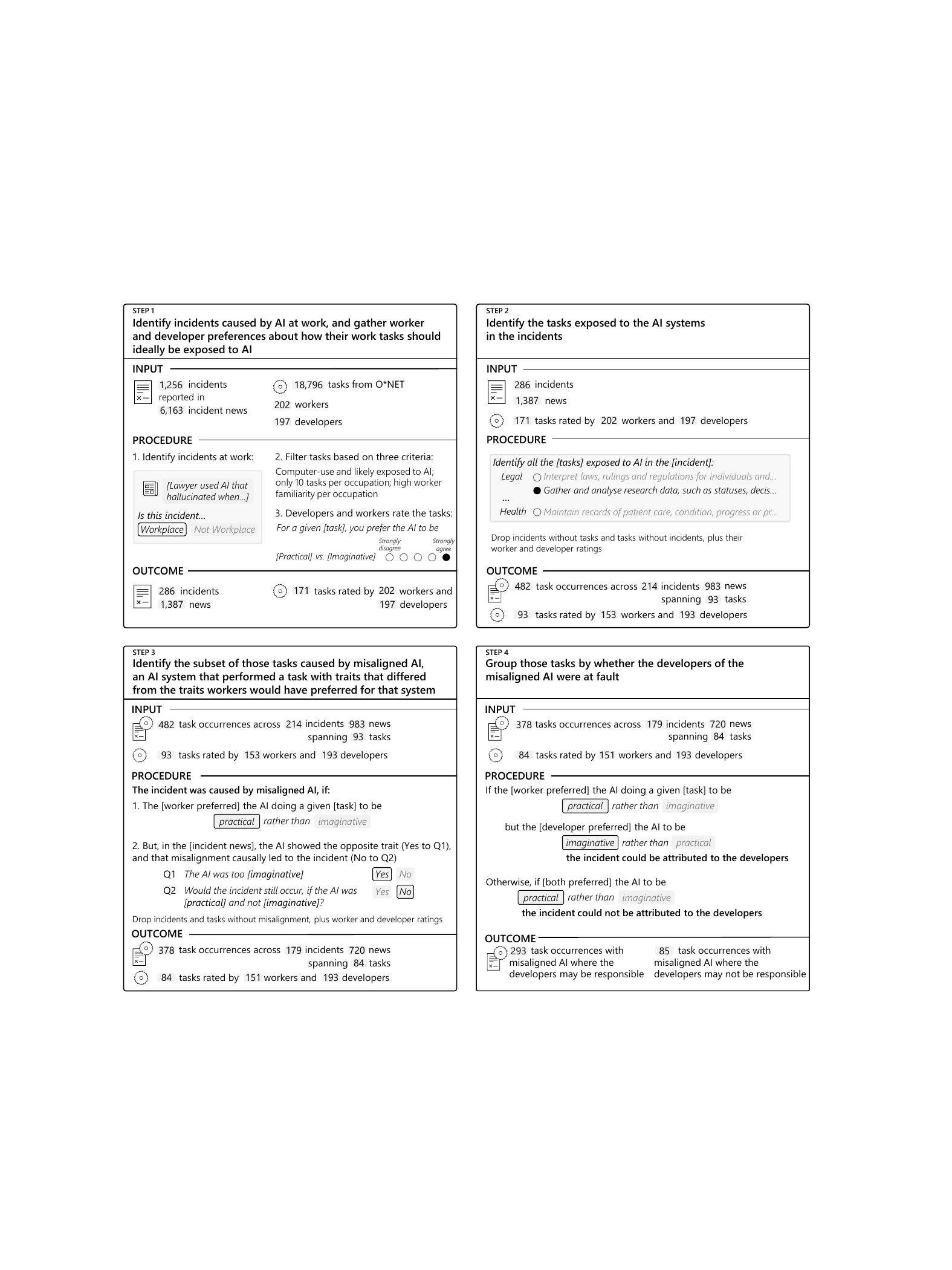} 
    \caption{\textbf{Overview of our research design.} We identify incidents caused by AI at work, and gather worker and developer preferences for how AI should be for a set of tasks from the O*NET (a standardized job task database) (Step 1); we identify the tasks exposed to the AI systems in the incidents (Step 2); we identify the subset of those tasks caused by misaligned AI (Step 3); and, we grouped those tasks by whether the developers of the misaligned AI were at fault or not (Step 4).}
    \label{fig:figure1}
\end{figure}

To answer our RQs, we followed four steps (Figure~\ref{fig:figure1}).
First, we identified incidents caused by AI at work, and gathered worker and developer preferences about how their work tasks should be exposed to AI (\S\ref{sec:methods_step1}).
Second, we identified which tasks were exposed to AI (\S\ref{sec:methods_step2}).
Third, we identified the subset of those tasks that were caused by misaligned AI (\S\ref{sec:methods_step3}).
Fourth, we grouped those tasks by whether the developers of the misaligned AI were at fault (\S\ref{sec:methods_step4}).

\subsection{Identifying Incidents Caused by AI at Work, and Gathering Workers' and Developers' Preferences About How Their Work Tasks Should Ideally Be Exposed to AI}\label{sec:methods_step1}

\noindent
\textbf{Collecting incidents caused by AI from an AI incident database.}
There are databases collating news involving AI incidents~\cite{turri_why_2023}. Out of all databases, we took the AI Incident Database (AIID) \cite{mcgregor_preventing_2021} for its broad coverage and wide use in prior work \cite{nedzhvetskaya_no_2024, Li_aies_2025, demiguel_incidents_2024, richards2025incidents}. Other platforms provide limited access to the news (e.g., paywall), and rely on automated collection with little human review. 
The AIID allows users to submit incidents supported by sources, mostly news, for editorial review \cite{mcgregor_preventing_2021}. 
The AIID hosts incidents with news covering the period from 2013 to 2025, with submissions increasing over time \cite{standford2025ai}. We collected all 1,256 incidents (reported by 6,163 news) up to November 2025.

\vspace{0.5em}
\noindent
\textbf{Identifying those incidents that occurred at work.}
We used an LLM approach to classify whether incidents occurred in the workplace (Step 1 in Figure~\ref{fig:figure1}).
We performed all classifications using the GPT-5 API \cite{openai2023gpt5}. 
We designed a prompt informed by prior work (Appendix~\ref{appendix:prompt_workplace}) \cite{shao_future_2025, demiguel_incidents_2024, loaiza_epoch_2024, brown2020language}, including a definition of workplace, workers, and work exposed to AI. 
We applied the prompt on a random set of 100 incidents and supporting news.  
To validate this classification and finalize our prompt, we performed three steps. First, two researchers annotated the same 100 incidents independently. We measured agreement using Cohen's kappa, a chance-corrected measure of inter-rater agreement~\cite{cohen1960coefficient,landis1977measurement}. The two researchers reached strong agreement with a Cohen's Kappa of 0.79. This human-to-human agreement was used as the reference level for the LLM annotation task.  Second, the researchers reached consensus on the annotations and compared those with the initial LLM annotations. This comparison yielded a Cohen's kappa of 0.66, generally considered substantial \cite{landis1977measurement}, but falling below our previously found reference level. Based on visual inspection, we determined the  main sources of error. Third, we revised the prompt by adding filtering criteria to exclude those  sources of error, and reclassified the same 100 incidents.
Cohen's kappa increased to 0.85, exceeding the reference level. 
Having finalized our prompt, we then applied it to all the incidents.  
In total, 286 incidents (22.7\%, reported by 1,387 news) were classified as having occurred at work.

\vspace{0.5em}
\noindent
\textbf{Collecting tasks and recruiting workers and developers.} 
To curate a representative set of tasks and recruit workers and developers, we drew on a user study that should satisfy two criteria.
First, the study had to examine specific task-level AI use within occupations. This returned two studies~\cite{ranjit_are_we,shao_future_2025}.
Second, it should have publicly available data and contactable participants.
We selected ~\cite{ranjit_are_we} as our primary framework as it met both criteria.

We performed two procedures drawing on the selected study
~\cite{ranjit_are_we} 
(Step 1 in Figure~\ref{fig:figure1}). 
First, we collected 18,796 tasks from the O*NET database \cite{onet2026}.
We also recruited the workers and developers from the study (Appendix~\ref{appendix:developers}).
We recruited 202 workers in Prolific and screened them for domain expertise. 
We also recruited 197 U.S.-based developers with AI expertise. All reported weekly AI use and held non-managerial engineering roles.
Second, we filtered the tasks following the study's criteria \cite{ranjit_are_we, shao_future_2025}.
We kept those likely exposed to AI, focusing on core, frequently performed, computer-based tasks (e.g., draft a report). 
The filter reduced the set to 2,078 tasks.
We further filtered the tasks to only include those that multiple workers rated as highly familiar. 
The filter returned 171 tasks across 12 sectors.

\vspace{0.5em}
\noindent
\textbf{Gathering worker and developer preferences of AI traits for a set of tasks.} 
To gather worker and developer preferences, we based our survey items on prior work. 
We set two criteria to select a framework on AI traits the AI should possess:
the framework had to be validated, and
the framework had to involve evaluation of traits in the workplace.
These filters returned three studies \cite{mckee2024warmth,dong_fears_2024,ranjit_are_we}.
We filtered out the study that explored workers' preferences for AI systems along two dimensions: warmth and competence \cite{mckee2024warmth},
as this two-part model was deemed too simple~\cite{dong_fears_2024}. 
Alternatively, Dong et al. found that people judge AI's job suitability based on eight pairs of traits, such as warm or imaginative \cite{dong_fears_2024}.
Ranjit et al. recently extended these eight pairs by adding four traits from responsible human-AI collaboration, such as explainable and open to challenge \cite{ranjit_are_we}. 
We used Ranjit at al.’s extended framework, which allowed for more nuance.
One limitation, however, is that the framework was developed with U.S.-based participants, and might be sensitive to the cultural context in which it was developed.
We used those 12 pairs of opposite AI traits (e.g., imaginative or practical) to survey the previously recruited 202 workers and 197 developers for the set of 171 tasks (Appendix~\ref{appendix:developers}). These rated which AI traits a system should have for tasks that are in their domain and are familiar with.
For example, participants from the legal sector highly familiar with drafting a case were asked ``For the task of drafting a case, to what extent should an AI system be imaginative and bring new ideas rather than stay practical and follow familiar approaches?''.
Workers and developers reported their preferences on a five-point Likert scale.
Preferences were reported on a five-point Likert scale, where ratings below 3 indicated a preference for one trait in a pair and ratings above 3 indicated a preference for the opposite trait. 
We defined AI trait misalignment as a gap of more than 0.5 points between worker and developer ratings, using this as a conservative threshold to focus on meaningful differences, in-line with the prior validated study \cite{ranjit_are_we}. 
For example, if workers rated a task at 2.4 (preferring practical) and developers rated it at 4.3 (preferring imaginative), the gap of 1.9 points exceeds the threshold and the task is classified as misaligned. 
In total, we collected preferences from 202 workers and 197 developers on 12 AI trait pairs for 171 tasks across 12 sectors.


\subsection{Identifying the Tasks Exposed to the AI Systems in the Incidents}\label{sec:methods_step2}
\vspace{0.5em}
\noindent
\textbf{Identifying tasks exposed to the AI systems in the incidents.} 
Since incidents could involve several tasks at the same time (for example, an AI tool used for both research and writing a report). We refer to each time a task appears in an incident as a \textit{task occurrence}. 
To identify which of the 171 tasks were exposed to AI in each of the 286 incidents (Step~2 in Figure~\ref{fig:figure1}), we extracted all task occurrences from the incident news using an LLM.
First, we designed a prompt that included the previously collected set of tasks, and asked whether the tasks were exposed to AI in the given incident (Appendix~\ref{appendix:prompt_tasks}). 
The LLM had to rate each task from 0 (not mentioned) to 3 (exposed to AI and involved in the incident). 
To ensure accuracy, we kept only tasks that scored 3.
We applied the prompt to a random sample of 100 incidents. 
To validate this classification and finalize our prompt, we performed three steps.
First, two researchers independently annotated whether they agreed with the LLM’s classifications.
Cohen’s kappa between the two researchers was 0.89. We used this as the reference level for the LLM annotation task. 
Second, the researchers reached consensus on the annotations and compared those with the LLM annotations, yielding a Cohen’s kappa of 0.76.
This is often considered substantial agreement \cite{landis1977measurement} but it did not reach our reference level. 
We found the main source of error was due to the LLM identifying tasks that were plausible but did not match the sector. 
Third, we refined the prompt to require that both the task and sector should match the incident description.
We applied the refined prompt to the same 100 incidents, and the Cohen’s kappa increased to 0.97, surpassing the reference level. 
We applied the finalised prompt to the full incident set. 
This yielded 482 task occurrences across 214 incidents (74.8\%), reported in 983 news articles, spanning 93 unique tasks (53.4\%). 

\vspace{0.5em}
\noindent
\textbf{Dropping incidents and tasks.}
In the process, we dropped 66 incidents (from 286 to 214) because none of their tasks were in our dataset, and 78 tasks (from 171 to 93) because they did not appear in any incident.
While this reduces the sample by nearly a quarter, keeping only tasks that appeared in incidents allowed us to compare directly the incidents with the survey.
This also affected the number of workers and developers in our analysis, as we kept only those who had rated at least one of the 93 remaining tasks.
The previous sample of 202 workers and 197 developers rating 171 tasks was reduced to 153 workers and 193 developers rating 93 tasks across 12 sectors.

\subsection{Identifying the Subset of Task Occurrences with AI Misalignment (RQ1)}\label{sec:methods_step3}

To identify which of the 482 task occurrences involved AI misalignment, we extracted the traits the system showed (Step 3 in Figure~\ref{fig:figure1}), and compared these traits with the previously collected workers’ preferred AI traits for a given task (Step 1 in Figure~\ref{fig:figure1}, \S\ref{sec:methods_step1}).
By \textit{misaligned AI}, we mean an AI system that performed a task with traits that differed from the traits workers would have preferred for that system.
The trait assignments are incident and AI-specific, ensuring we capture the variation across different AI systems, and their own failure modes.

\vspace{0.5em}
\noindent
\textbf{Defining a prompt to identify which tasks the AI did not perform in line with worker preferences (that is, misaligned AI tasks). } 
We designed a prompt to assess AI misalignment for each task occurrence (Appendix~\ref{appendix:prompt_traits}).
The prompt included the AI incident description, its supporting news, and task description. 
The prompt was structured in two parts.
First, the model analysed the incident news and identified textual evidence of whether the AI exhibited any of the 12 AI traits while performing the task.
Second, for identified traits, the model assessed whether misalignment contributed to the incident.
We used a counterfactual definition of causality: misalignment was causal if the incident would not have occurred without it.
Two questions (Q) operationalised this.
The first (Q1) assessed whether the AI showed the opposite trait (e.g., “The AI was too \emph{imaginative}”). 
The second (Q2) assessed whether that misalignment causally led to the incident (e.g., “Would the incident still occur if the AI was \emph{practical} and not \emph{imaginative?}”). 
We classified an incident as caused by misaligned AI when Q1 returned ‘Yes’, and Q2 returned ‘No’, and the worker indeed preferred the opposite trait.
This definition does not rule out other contributing factors outside our framework.

\vspace{0.5em}
\noindent
\textbf{Running the prompt to identify the tasks caused by AI misalignment and validating the results.} We ran the prompt on 100 randomly selected incidents.
To validate the accuracy of this classification, two researchers independently reviewed the LLM outputs and annotated whether they agreed with the trait assignments. 
Human-to-human agreement was a Cohen's kappa of 0.87, set as the reference level for this annotation task. 
The authors reached consensus on their disagreed annotations. We compared those human annotations against the LLM prior annotations and reached the reference level (Cohen’s kappa = 0.90), indicating reliable extraction. 
We applied the prompt to all the remaining incidents.
This returned 378 task occurrences with misaligned AI across 179 incidents reported by 790 news, spanning 84 unique tasks. 
As some tasks showed no misalignment across incidents, we dropped 9 tasks (from 93 to 84) and 35 incidents (from 214 to 179), which also reduced our sample to 151 workers and 193 developers who had rated at least one of the 84 remaining tasks.
To explore these misalignments, we read in-depth 10\% of the incidents. We employed close reading, which is a qualitative method that consists in reading a document line by line in detail, paying attention to nuances that uncover underlying meaning \cite{janicke2015close, gutierrez_tool_2025}. We randomly selected two sources for each incident, as the number of news sources varied.

\subsection{Grouping Those Tasks by Whether the Developers of the Misaligned AI Were at Fault (RQ2)}\label{sec:methods_step4}

We grouped the 378 task occurrences with misaligned AI into two groups: those that could be attributed to the developers, or those that could be attributed to other causes (Step 5 in Figure~\ref{fig:figure1}). 
We used the previously collected 153 workers' and 193 developers' preferences of AI traits for 84 tasks, and their misalignment (from Step 3).

\vspace{0.5em}
\noindent
\textbf{Establishing the rule to determine whether developers of misaligned AI could be considered responsible or not.}
We attributed incidents to developers when the AI showed traits that the workers did not want but the developers had deliberately designed.
For each of the task occurrences, we established the following grouping rule.
If workers preferred the AI to show one trait while performing a task, but developers deliberately designed the opposite trait, any resulting incident could be attributed to the developers. 
For example, workers may have wanted the AI to be practical rather than imaginative when drafting a report, whereas developers may have designed it to be imaginative.
In this case, the fault lies in the design stage.
Otherwise, if both the worker and the developer preferred the AI to show the same trait (e.g., be practical rather than imaginative) and this resulted in an incident,
then the incident could not be attributed to the developer.
In this case, the cause is harder to trace.

\vspace{0.5em}
\noindent
\textbf{Applying the grouping rule to group the tasks.}
Applying the previously defined rule resulted in two groups: 293 task occurrences where the developers may be responsible, and 85 task occurrences where the developers may not be responsible.
In line with the previous section~(\S\ref{sec:methods_step3}), we performed close reading of the documented incidents and their news sources to better understand the context of the incidents.

\section{Results}\label{sec:results}

\subsection{Prevalence and Patterns of Misaligned AI Causing Incidents in the Workplace (RQ1)}\label{sec:results_rq1}
To answer RQ1, we examined 214 workplace incidents and their corresponding 482 task occurrences, identifying those caused by misaligned AI (Step 3 in Figure~\ref{fig:figure1}). 

\begin{figure}[t]
    \centering
    \includegraphics[width=0.95\columnwidth]{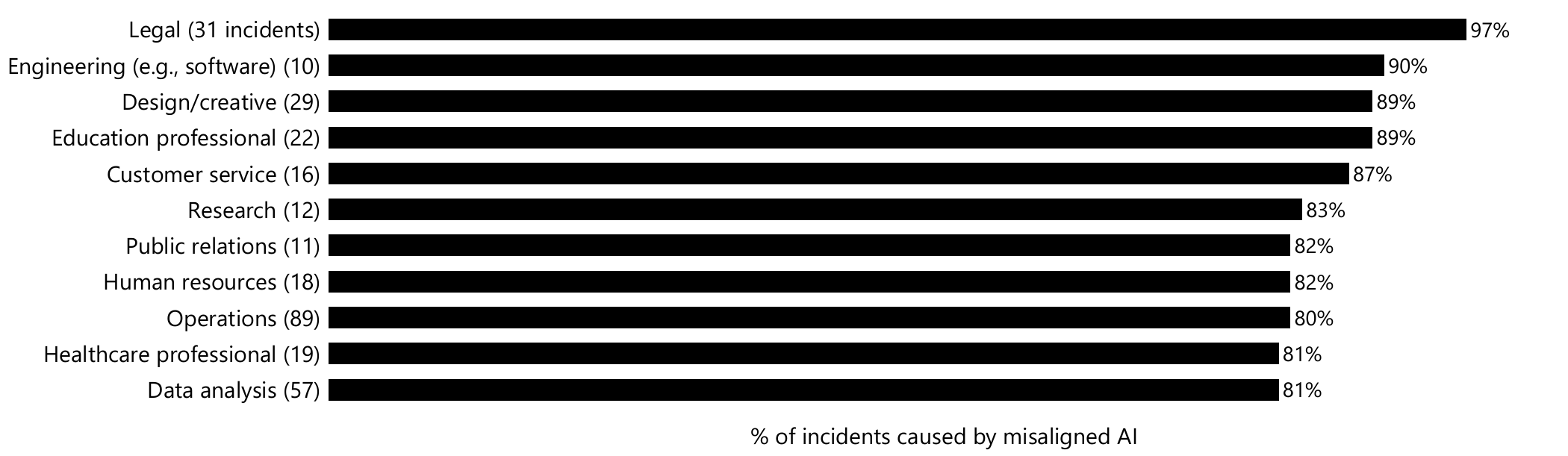}
    \caption{\textbf{
    Percentage of incidents caused by misaligned AI across sectors.} 
    By misaligned AI, we mean an AI system that performed a task with traits that differed from the traits workers would have preferred for that system.
    We count an incident as caused by misaligned AI, if at least one task occurrence in the incident was misaligned and this contributed to the incident.
    The bars represent the percentage of incidents caused by misaligned AI in a sector, with the total number of incidents per sector given in parentheses. 
    The majority of AI incidents are consistently caused by misaligned AI across sectors, predominantly in the legal, creative, education, and, interestingly, engineering sectors. The reasons for misalignments are explored in Figure~\ref{fig:figure4}.}
    \label{fig:figure2}
\end{figure}

\vspace{0.5em}
\noindent
\textbf{Misaligned AI plays a major role in causing incidents in the workplace.} 
RQ1 asked `To what extent are workplace AI incidents caused by misaligned AI?'
By misaligned AI, we mean an AI system that performed a task with traits that differed from the traits workers would have preferred for that system.
Worryingly, we found an overwhelming majority of the incidents in the workplace were caused by misaligned AI: 83.6\% (179 of the 214 workplace AI incidents).
These range from 81\% for data analysis to 97\% of the incidents for the legal sector (Figure~\ref{fig:figure2}).
The legal sector has the highest share (97\%), followed by engineering (e.g., software) (90\%), and the creative and educational sectors (both 89\%). 

\begin{figure}[t]
    \centering
    \includegraphics[width=0.70\columnwidth]{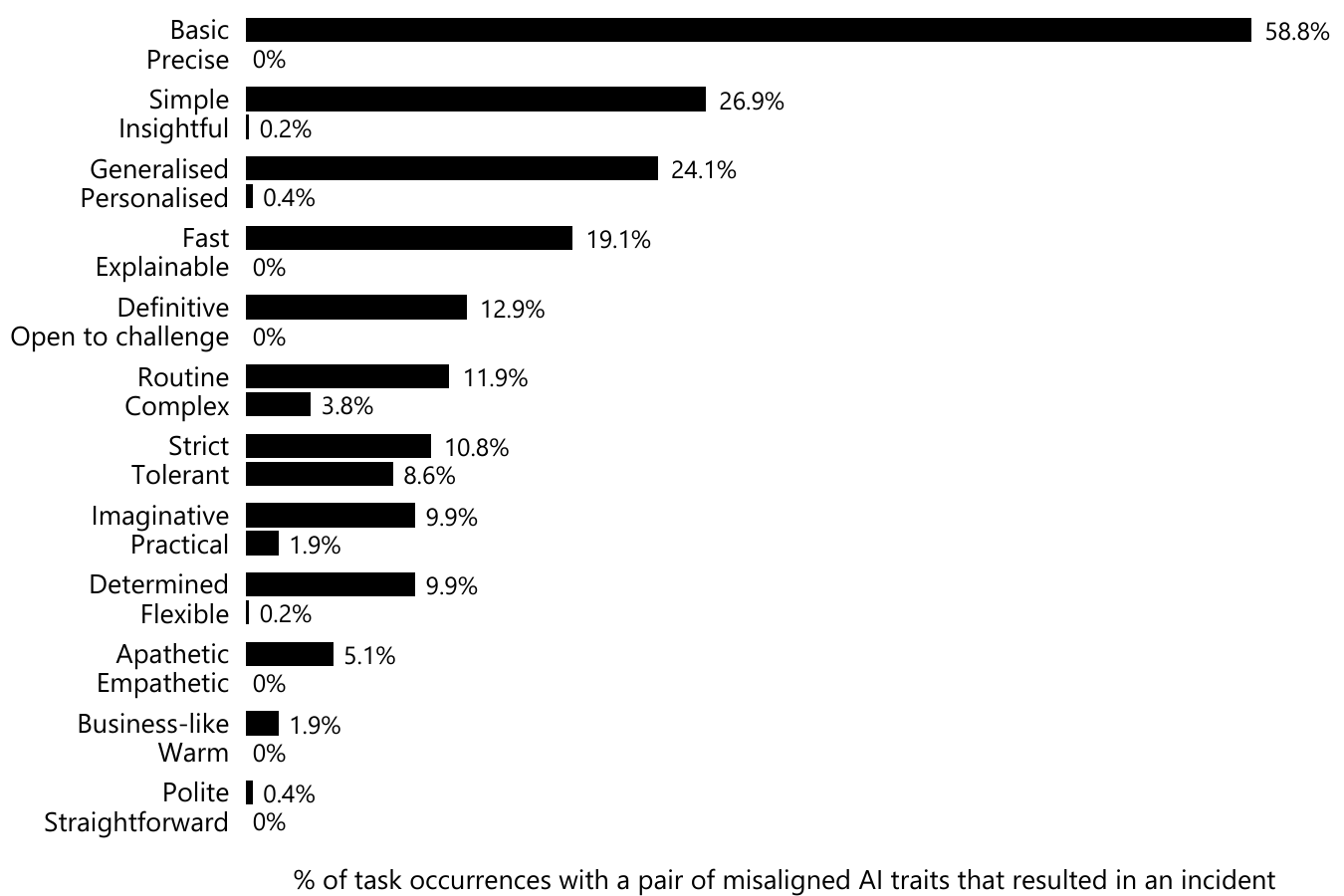}
    \caption{\textbf{
    Percentage of task occurrences exhibiting a given misaligned AI trait, out of all task occurrences with a pair of misaligned AI traits that resulted in an incident.} 
    A task occurrence is counted each time a task from our set of 93 tasks appears in one of the 214 incidents; since an incident may involve multiple tasks, the same task can be counted more than once.
    For each pair of AI traits, the two bars show the percentage of tasks where the AI displayed one trait (for example, basic or precise), while workers preferred the opposite trait, and that resulted in an incident.  
    For 58.8\% of task occurrences with AI misalignment, the AI provided \emph{basic} responses rather than the \emph{precise} answers workers expected.
    A task occurrence may involve AI misalignment on multiple pairs of traits. 
    The traits linked to the most incidents were caused by AI systems that were too basic, too simple, too impersonal, too oversimplified, and too fast. Results are mostly asymmetric, except for the strict \emph{vs.} tolerant pair.}
    \label{fig:figure3}
\end{figure}

\vspace{0.5em}
\noindent
\textbf{In most cases, workers want systems that are precise, insightful, or personal, but instead
receive systems that are basic, simple, or general.}
Globally, we found that the pair of misaligned AI traits that most frequently caused incidents involved the AI being \emph{basic} but the worker needing it to be \emph{precise} (across 58.8\% of task occurrences) (first row in Figure~\ref{fig:figure3}). 
This was followed by the AI being too simple, while the worker needed it to be insightful (26.9\%), and the AI being general, and treating everyone similarly, but the worker needed it to be personalised and adjust based on the individual (24.1\%). 

The pairs of traits in misaligned AI that cause incidents tend to show an asymmetry (Figure~\ref{fig:figure3}). 
This is, given a pair of AI traits, incidents mostly occur when the AI shows one trait (such as basic) while workers consistently preferred the opposite (such as precise), but rarely the reverse.
This does not mean the reverse misalignment cannot exist, but, when it does, it seems less likely to result in an incident. 
There are some exceptions for some pairs of traits, strict \emph{vs.} tolerant (10.8\% \emph{vs.} 8.6\%), routine \emph{vs.} complex (11.9\% \emph{vs.} 3.8\%), and, partially, imaginative \emph{vs.} practical (9.9\% \emph{vs.} 1.9\%).

\begin{figure}[t]
    \centering
    \includegraphics[width=1\columnwidth]{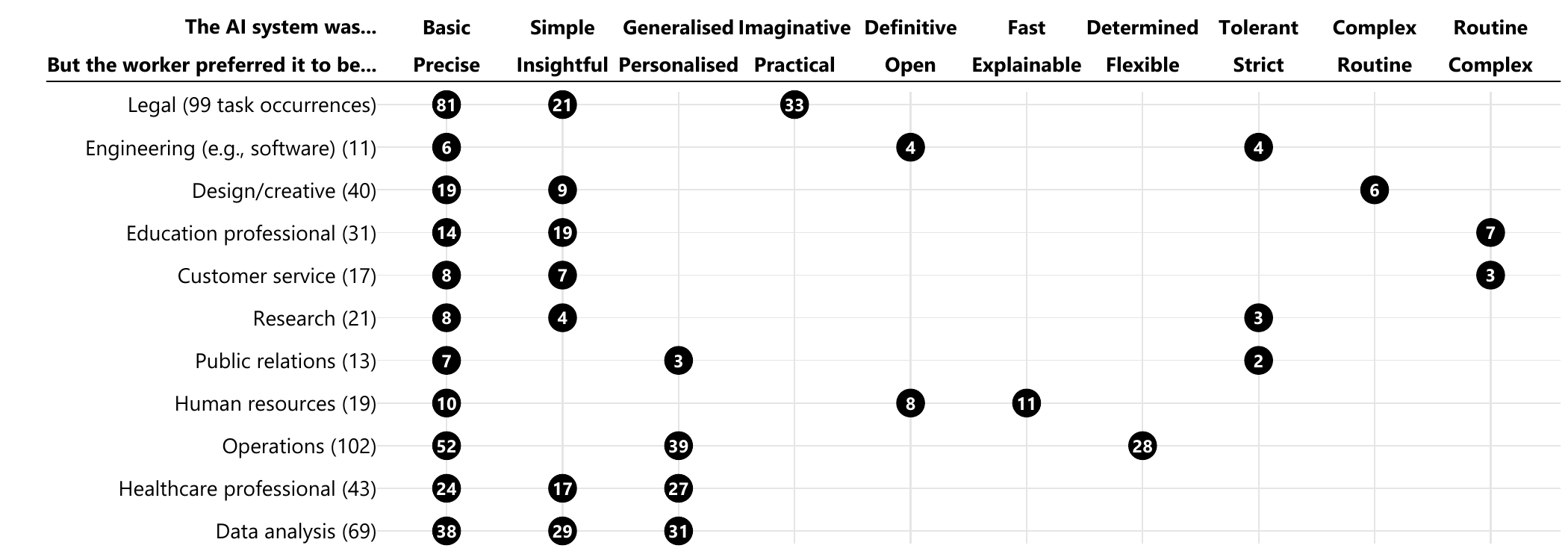}
    \caption{\textbf{
    Top three most frequent reasons for misalignment in each sector.}  
    By misaligned AI, we mean an AI system that performed a task with traits that differed from the traits workers would have preferred for that system (first two rows in the graph).
    The numbers in the black circles represent the number of the sector's
    task occurrences where the AI exhibited a given trait and the worker preferred the opposite trait, 
    out of all the sector's task occurrences with AI misalignment (in parentheses). 
    A task occurrence may involve AI misalignment on multiple pairs of traits. 
    In the legal sector, imaginative AI was responsible for incidents totalling 33 task occurrences, while in human resources, fast AI accounted for 11.} 
    \label{fig:figure4}
\end{figure}

Noticeably, we found some traits did not cause any incidents (Figure~\ref{fig:figure3}). 
No incidents involved an empathetic AI (0\%), focused on addressing human needs and emotions, when the worker just wanted an AI doing data handling; neither a warm AI (0\%), showing care, when the worker preferred it to remain business-like. 
Further, no incidents happened because the AI was explainable (0\%), making decisions that are easy for people to understand, but the worker needed a fast AI, with automatic decisions without explanations. Finally, incidents did not happen because an AI that was too open to challenge, but the worker preferred a definitive AI. 
It is also remarkable that the pair of traits of AI being too polite, even if the system is not fully honest, or too straightforward, almost none caused any incident in any direction (0.4\%/0\%). 
Yet, incidents involving AI being too straightforward have been documented outside the workplace, such as in mental health contexts, where overly direct AI responses have caused harm; we reflect on this in the discussion (\S\ref{sec:discussion}).

\begin{figure*}[t]
    \centering
    \includegraphics[width=\textwidth]{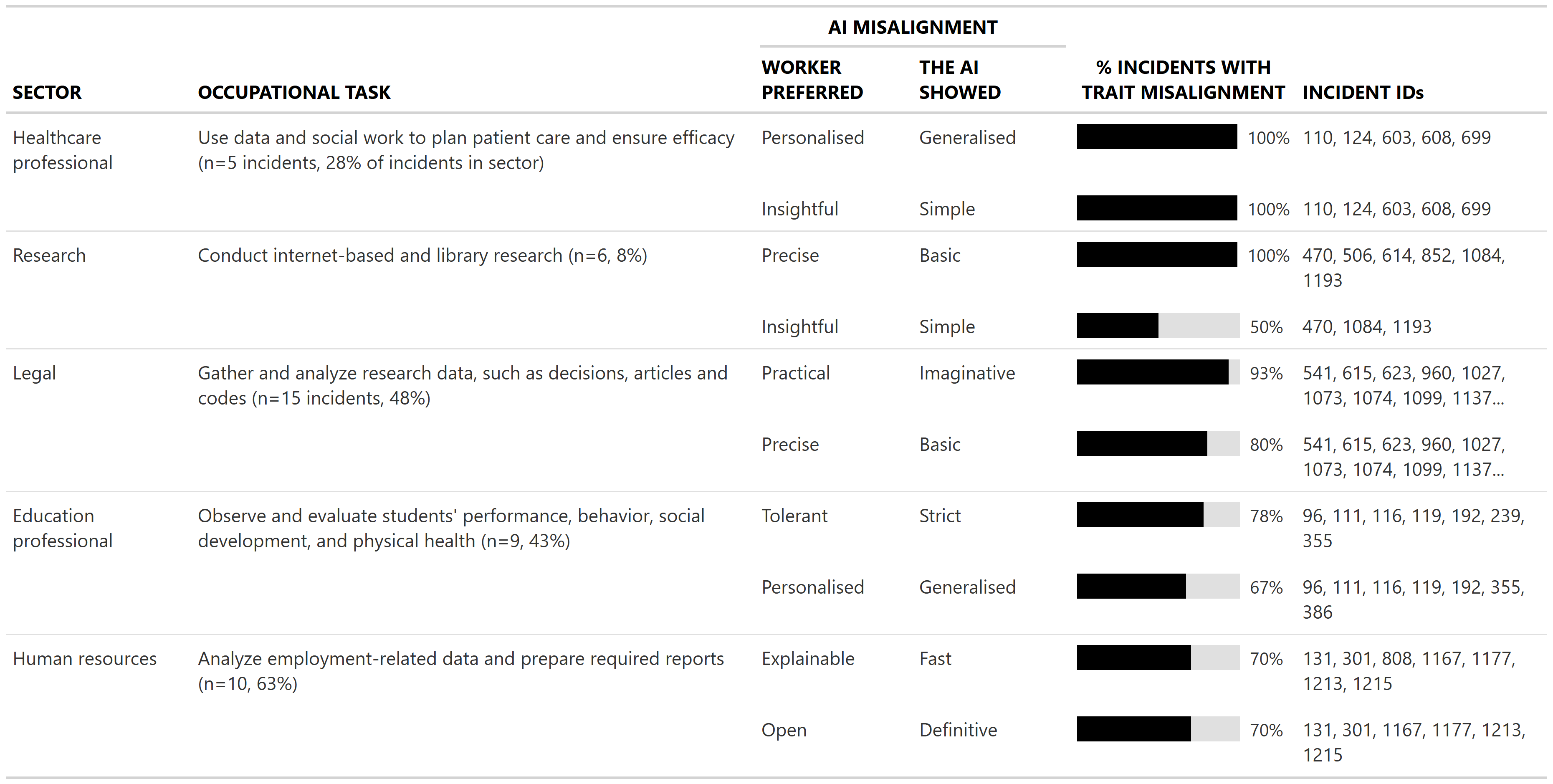}
    \caption{\textbf{
    The tasks with a higher fraction of incidents caused by misaligned AI.} 
    For each task, we show the two most prevalent traits where an AI system  performed a task with a trait that differed from the trait the workers would have preferred for that system.
    We report the percentage of incidents associated with a given task in which the AI showed a given trait and was misaligned, out of all incidents linked to the task (we provide incident IDs). An incident can involve the AI showing many misaligned traits for a given task. We limited the analysis to tasks across at least five incidents. 
    In general, when AI systems are too basic, incidents may well occur.
    }
    \label{fig:figure5}
\end{figure*}

We then honed in the most frequent reasons for misalignment in each sector, and we found key differences. 
We explored the top three pairs of misaligned traits across task occurrences that resulted in an incident (Figure \ref{fig:figure5}). 
This analysis revealed that the sectors of healthcare professional and data analysis coincide with the global top misalignments of basic, simple, and generalised systems, but others do not. 
Sectors in which the top misalignments do not coincide with the global ones are the legal sector, where imaginative AI instead of being practical creatively making up references (33 task occurrences); the human resources sector with AI systems used to recruit being too fast and definitive, but not explainable (11); the educational sector, with a routine AI used with students, when it should be more complex (6) (Figure \ref{fig:figure4}).

\begin{figure*}[t]
    \centering
    \includegraphics[width=1\columnwidth]{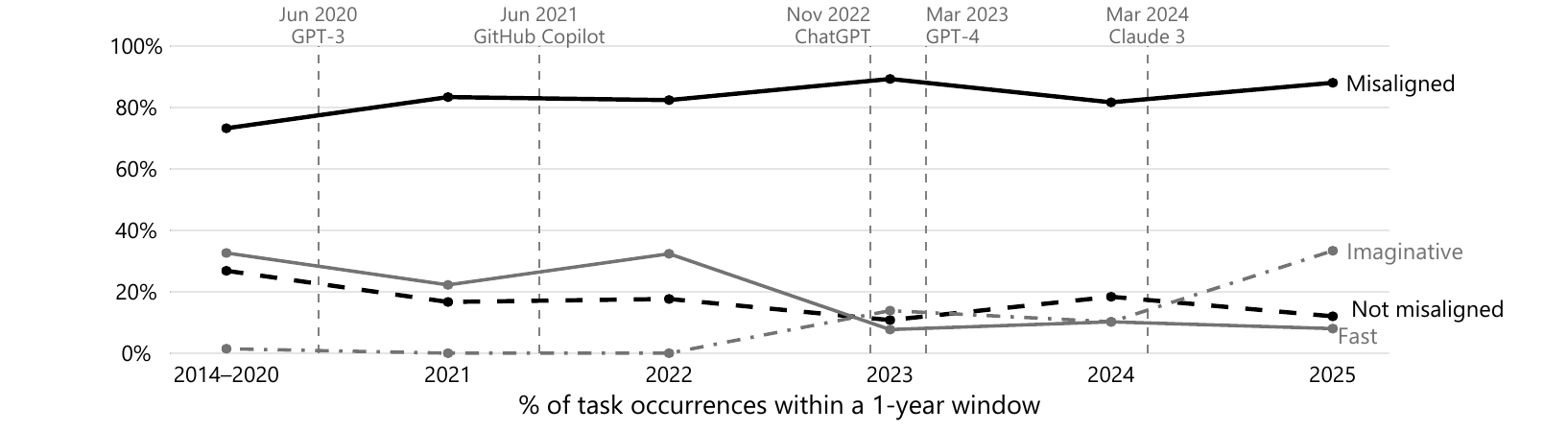}
    \caption{
    \textbf{Percentage of task occurrences involving incidents of misaligned AI (solid line), or incidents of not misaligned AI (dashed line) over time.}
    Each data point represents the percentage of task occurrences classified as specified in the label within a year. 
    We merged 2014-2020 as there were significantly fewer data points.
    Vertical lines mark milestones in AI research and deployment for context \cite{standford2025ai}.
    Misalignment consistently remained the dominant cause of incidents, although the frequency of misalignments involving specific traits increased or decreased over time: fast AI used to cause incidents, yet, since 2022, imaginative AI started to do so (Appendix~\ref{appendix:chronological} reports the results for all traits).}
    \label{fig:figure8}
\end{figure*}

Finally, we examined if misalignment changed over time (Figure~\ref{fig:figure8}). 
Misalignment consistently dominates task occurrences across the observed period (2014-2025), suggesting misalignment with workers persists despite rapid AI advances. 
Yet, the trends diverge at the trait-level. 
Fast AI decreasingly causes incidents over time, possibly reflecting advances in explainable AI that have made ML systems more predictable (Appendix~\ref{appendix:developers}). 
Conversely, imaginative AI increasingly causes incidents post-2022, likely driven by generative AI systems producing unreliable outputs. 
This suggests that as one form of misalignment declines, another emerges, keeping the total rate consistently high.

Now we turn to the close reading of incidents with the most misaligned tasks (Figure~\ref{fig:figure5}), starting with the task `Gather and analyse research data, such as decisions, articles and codes' in the legal sector,
(third row)."
The main trait misalignment happened when the lawyers needed a practical system (as indicated in our survey), but the AI system was imaginative (in 93\% of the incidents with the task). This was followed by the case when the worker needed a precise system that in turn was basic (80\%). 
Across incidents, the AI tool developed for the legal sector was making up fictitious references, and even when summarising the results it was not being precise enough. For example, a South African legal team used Legal Genius AI, an AI tool, and discovered this misalignment when the system generated non-existent case law in an urgent court filing, despite allegedly being trained on South African legal precedents (ID 1139) \cite{cdh2025aicitations}. 

Another example comes from the human resources (HR) sector, involving the task `Analyse employment-related data and prepare required reports', which occurs in 63\% of the sector's incidents (\emph{n}=10) (fourth row in Figure~\ref{fig:figure5}).
In our study, HR workers indicated that, when AI augments this task, they need the system to be explainable and open. Yet in the incidents we analysed, AI systems were designed to be the opposite: fast (causing 70\% of incidents with the task) and definitive instead (70\%), misaligned with what workers needed to perform their task. 
Amazon's algorithmic management system for Flex drivers illustrates this misalignment can cause harm (ID 111). 
The system automated firing decisions based primarily on punctuality metrics, with algorithms that ``scan the gusher of incoming data for performance patterns,'' while ``human feedback is rare'' \cite{soper2021fired}. 
Rather than supporting HR workers with an explainable system, this fast design model prioritized measurable metrics and simple rules for efficiency.
As a result, the harm fell on the drivers that HR workers were supposed to oversee. 
The misalignment resulted in wrongful terminations, with drivers penalized for traffic, weather, or route complexity factors the system's definitive logic could not accommodate.

\subsection{Developers’ Design Decisions Resulting in AI Incidents in the Workplace (RQ2)}\label{sec:results_rq2}
To answer RQ2, we examined 179 incidents and 378 tasks occurrences, and identified whether the developers could be considered responsible for the misalignment, or whether it was due to other causes (Step 4 in Figure~\ref{fig:figure1}). 

\begin{figure}[t]
    \centering
    \includegraphics[width=0.9\columnwidth]{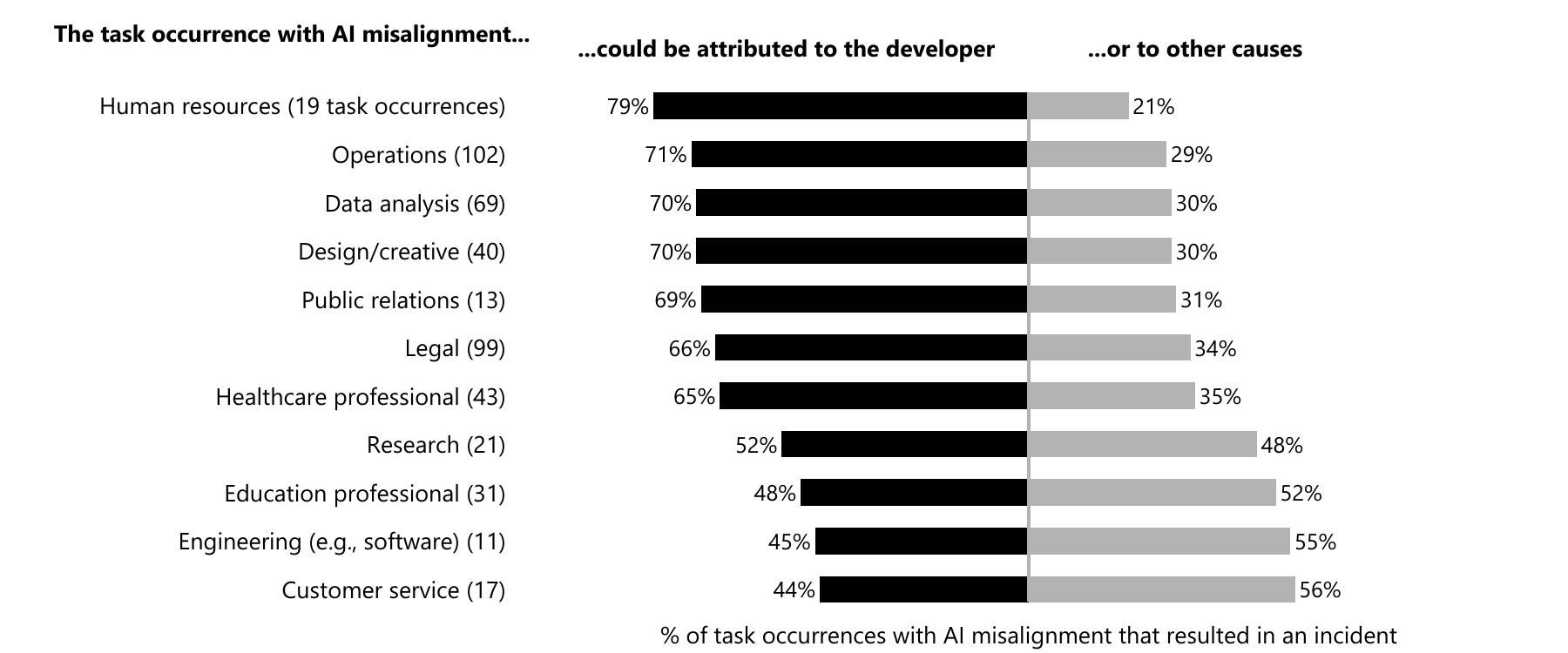}
    \caption{
    \textbf{
  Percentage of task occurrences in incidents with AI misalignment attributable to developers (black bar) or to other causes (gray bar), out of all task occurrences with AI misalignment within each sector (total in parentheses).} 
  We attribute the responsibility of a task occurrence with AI misalignment to the developer, if the developer designed the AI to perform a task with at least one trait that differed from the traits workers would have preferred for that system.
  Developers’ design intentions may account for more than half of the task occurrences with AI misalignment that resulted in incidents (74\%), mainly in working-facing sectors such as human resources. The reasons for misalignments are explored in Figure~\ref{fig:figure7}.}
    \label{fig:figure6}
\end{figure}

\begin{figure}[t]
    \centering
    \includegraphics[width=1\columnwidth]{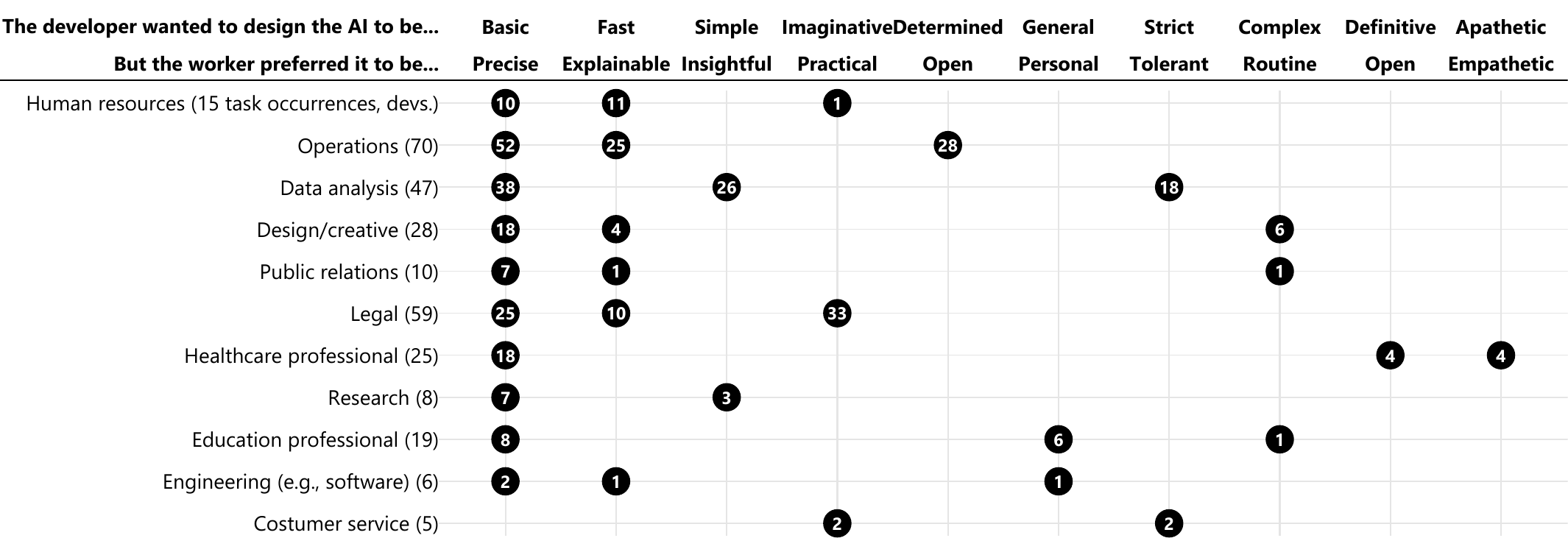}
    \caption{
    \textbf{
    Top three most frequent misalignment attributable to developers in each sector.} 
    We attribute the responsibility of a task occurrence with AI misalignment to the developer, when the developer designed the AI system to perform a task with traits that differed from the traits workers would have preferred for that system (first two rows in the figure).
    The numbers in the black circles represent the number of the sector's
    task occurrences where the developer designed for a given trait and the worker preferred the opposite trait, 
    out of all the sector's task occurrences with AI misalignment that could be attributed to the developer (in parentheses). 
    A task occurrence may involve AI misalignment on multiple pairs of traits. 
    Developers are most misaligned with workers on traits that prioritize efficiency and speed (basic and fast).
    }
    \label{fig:figure7}
\end{figure}

\vspace{0.5em}
\noindent
\textbf{Developers’ may be responsible for more than half of the tasks occurrences with misaligned AI, mainly in the human resources and public relations sectors.} 
RQ2 asked `When trait misalignment results in incidents, how often could it be attributed to developers \emph{vs.} other causes?'
We found 73.6\% of the task occurrences with AI misalignment could be attributed to the developer. 
This is, when the developers design an AI system to perform a task with traits that differ from the traits that workers would have preferred for the system, and that resulted in incidents.
The remaining 26.4\% tasks involved traits where both developers and workers agreed on all the traits that the AI should have for a given task. 
In this case, the misaligned AI occurred due to other reasons, which also resulted in an incident. 

Figure~\ref{fig:figure6} shows the percentage of task occurrences with misaligned AI that could be attributed to developers or to other causes, out of all task occurrences with AI misalignment within in each sector. 
We find some sectors where misalignment at the design stage largely translates into workplace incidents, with human resources and operations accounting for 79\% and 71\% of the tasks occurrences with AI misalignment. 
In the sectors of costumer service and engineering, more incidents caused by misalignment were attributed to other causes than to developers, who accounted for 44\% and 45\% of the tasks occurrences. Notably, in some sectors the attribution of misalignment is more balanced between developers and other causes, like in research and education.

\vspace{0.5em}
\noindent 
\textbf{Developers differ most from workers on traits that prioritize efficiency and speed.}
We found the most misaligned traits across sectors are basic, fast, imaginative, determined, generalised, strict, tolerant, complex, and apathetic, but workers preferred precise, explainable, practical, open, personalised, tolerant, strict, routine, and empathetic AI (Figure~\ref{fig:figure7}).
All those pairs of traits also appeared in the most frequent trait misalignment (Figure~\ref{fig:figure4}), but apathetic. 
These traits range from one sector, e.g., apathetic AI healthcare in, to ten sectors, e.g., basic AI. 

A pattern appears: systems were designed for efficiency-driven traits (basic or fast) while workers needed traits for the contextual complexity and stakes of their tasks. 
One case previously discussed (\S\ref{sec:results_rq1}) involves an HR task for the analysis of employment records, where developers rated fast as more important than workers, who preferred explainability. 
This misalignment resulted in incidents involving Amazon’s Flex driver systems (ID 111, ID 116). The AI automatically processed workers' performance data and output termination decisions with minimal to no human oversight. 
Statements from developers explain misaligned design choices. For example, in ID 111, an engineer said ``Inside Amazon, the Flex program is considered a great success, whose benefits far outweigh the collateral damage'' \cite{soper2021fired} suggesting that speed and automation were prioritized over interpretability.

A similar pattern emerges in incidents involving AI agents supervising technical workers. In our study, engineers preferred open over definitive,
precise over basic,
and complex over simple.
In ID 194, an AI system made field deployment decisions that could not be overridden, despite producing false positives. Engineers noted that ``the automation team was unable to turn off the bot'' in cases that ``only a human operator could understand'' \cite{tamboli2020lesson}. These failures reflect design choices that prioritised definitive and simplified decision-making where workers needed agency and flexibility.

\section{Discussion}\label{sec:discussion}
In this section, we contextualise our findings in relation to prior literature~(\S~\ref{sec:related}), outline the implications of our research~(\S~\ref{sec:implications}), and list the main limitations, while offering directions for future work~(\S~\ref{sec:limitations}).

\subsection{Overview of Our Findings in Relation to Prior Literature}\label{sec:related}
\vspace{0.5em}
\noindent
\textbf{Consistency with prior literature.} We discuss two findings that are consistent with the literature.
First, our findings empirically demonstrate that trait (mis)alignment is contextual and task-specific, rather than universal. 
While some of the AI traits affect all sectors similarly (like basic AI, simple AI, and general AI) other traits matter more for some sectors (Figure~\ref{fig:figure4}).  
This resonates with Suchman's concept of \textit{situated action} \cite{suchman2007human}, where an interaction is contingent on the context.
For instance, we found that imaginative AI is problematic in the legal sector, and apathetic AI is so in the healthcare sector (Figure~\ref{fig:figure7}). 
A recent study found that people feared the same misaligned traits we identified such as excessive imagination when augmenting a lawyer's tasks, or not addressing human needs in medical tasks \cite{fears_ai_dong_2024}.

Second, by exploring where trait misalignment originates in the AI lifecycle, we demonstrate that many misalignments already exist in developer design. 
We found developers are most misaligned with the worker-facing sector of HR~(Figure~\ref{fig:figure7}). 
Prior work shows developers often lack familiarity with the sectors they build for, resulting in a `failure of imagination'~\cite{harvey_dont_2025, kawakami2026ai,boyarskaya2020overcoming}.
Developer demographics partly explain this: in our study, they were mostly U.S.-based, computer science-trained, and male, often misaligned with responsible AI practices \cite{olson2025speaks,jakesch_how_2022}.
Yet, our close reading revealed that developers' misalignment also reflected organizational priorities, such as speed when workers needed explainability~(\S\ref{sec:results_rq2}). 
Developers' decisions are not solely their own choices, but are shaped by the wider environment in which they work, prioritising employers over worker needs~\cite{widder2023s,gorucu2025individual,rakova2021responsible}. 


\vspace{0.5em}
\noindent
\textbf{Novelty compared to prior literature.}
We discuss two findings that contribute to the literature.
First, our analysis demonstrated that AI trait misalignment with worker needs plays a key role in workplace incidents. 
Worryingly, it accounts for 83.4\% of the incidents in our dataset (Figure~\ref{fig:figure2}).
Prior work had found negative effects of misaligned AI. 
For example, clinicians struggled to uptake LLMs to summarise their clinical notes, as the AI was too strict for their open-ended, `messy' notes \cite{kupferschmidt_write_2025}.
However, no study had shown misalignment leads to broader incidents in the workplace. 
We provide the first macro-level assessment of trait misalignment across 12 sectors, grounded in real-world incidents.

Second, we developed a novel framework to analyse workplace AI incidents through the lenses of HCI, enabling researchers to systematically link AI failures to design choices and prevent them at design stage~\cite{kawakami2026ai}. 
This approach is well-established in safety-critical fields such as aviation and healthcare. 
Methods in aviation include gathering detailed incident reports, and testing how user interfaces behave under edge-case conditions~\cite{turri_why_2023}.
Our framework integrates user experience into incident analysis, which often focuses primarily on system outputs \cite{gillespie2026ai,rauh2024gaps,rismani2025measuring}.




\subsection{Implications}\label{sec:implications}
\vspace{0.5em}
\noindent
\textbf{Trait misalignment is a major risk factor, and should be incorporated into risk assessments.}
Responsible AI risk assessments should incorporate HCI indicators, like AI trait misalignment, to anticipate deployment contexts and worker needs, following HCI design guidelines accounting for user mental models~\cite{weisz_design_2024,norman2013design,amershi_guidelines_2019}.
For example, while the EU AI Act mandates risk assessments only for high-risk systems \cite{golpayegani2023high}, lower-risk workplace AI may still cause harm through trait misalignment \cite{bogucka_atlas_2024}, a concern the Act does not currently address.

\vspace{0.5em}
\noindent
\textbf{Preventing misalignment requires structural interventions at the design stage.}
Developer teams should incorporate participatory approaches from the earliest stages of development. 
These approaches can draw from HCI and CSCW methodologies, including co-design workshops, and iterative feedback with workers~\cite{sadeghian2025workai, madaio2020chi}.
Yet, to be effective, interventions must be structural by altering the cultural context \cite{blankenship2006structural}. As such, interventions should address the drivers of developers' misalignment at three levels: micro, meso, and macro.
At the micro level, they should tackle developers' assumptions stemming from their technical training, and intersectional experience~\cite{olson2025speaks,birhane2022forgotten,gebru_datasheets_2021,d2023data}. 
At the meso level, they should address the companies where developers work, and how they envision innovation and for \emph{whom} (e.g., do edtech startups genuinely centre school teachers?) \cite{harvey_dont_2025,karizat2024patent,wajcman2024rebalancing}.
At the macro level, they should challenge how geopolitical or economic pressures push agendas when discussing worker needs (e.g., framing automation as a national imperative for global competition)~\cite{elish2018situating, elish_moral_2019}.
Without meaningful structural interventions, we risk \textit{participation washing}~\cite{sloane2022participation}.


\vspace{0.5em}
\noindent
\textbf{There are additional sources of failure that the literature should still study.}
We observed cases where developers were aligned, yet AI failed to exhibit the traits required to perform the task. 
AI capabilities may have limitations for alignment ~\cite{holstein2019improving,keyesML}.
For example, research has shown how trying to personalise AI for fair assessments may not work in the real-world, e.g., \textit{fairML}~\cite{jorgensen2023not,lighthouse2025}.
Despite the AGI hype on AI capabilities, future work should explore its limitations.
%

\subsection{Limitations and Future Work}\label{sec:limitations}

Our study has four main limitations.
First, our AIID incident data mostly capture newsworthy and US-focused harms \cite{demiguel_incidents_2024, richards2025incidents}, potentially missing everyday worker experiences. 
For example, while polite AI rarely caused incidents (0.4\%) (Figure~\ref{fig:figure3}), caution is due.
Sycophantic AI, too agreeable at the expense of accuracy~\cite{sharma2023towards}, causes severe harms in mental health \cite{cheng2025social}. 
Our finding could be explained for two reasons. 
First, our data is not exhaustive, and polite AI incidents may not be newsworthy.
Second, these failed interactions take time to surface. 
Future work should explore the implications of polite AI at work, with longitudinal studies or ethnographies.

Second, our task-level approach does not fully operationalise a job or an AI product \cite{autor_task_2013,wajcman2011constant}. 
Future work could consider additional dimensions of jobs, such as informal activities, contextual organisational fit, workers' skills, and AI literacy; and additional dimensions of AI products, such as AI type, and company~\cite{autor_task_2013}.

Third, our counterfactual approach (traits are causal if their absence would not result in incidents) cannot definitively establish causation as we would in controlled experiments. While we rely on analyst judgment informed by incident descriptions, validated with high inter-rater reliability (Cohen's Kappa=0.89), further work could explore other contextual causes of the incident.

Our last limitation concerns the user study compatibility. Our incident analysis yields classifications on whether the trait caused incident (yes/no), while the user study we draw upon measured misalignment on a scale. We bridged this through an established threshold of 0.5 Likert point difference between developers and workers~\cite{ranjit_are_we}.

\section{Conclusion}
We systematically analyzed 214 workplace AI incidents through 12 widely-used pairs of psychological traits to assess AI misalignment.  
Worryingly, we found an overwhelming majority of incidents (83\%) involve trait misalignment.
We also found that 74\% of the misaligned tasks already existed in the developers design choices. 
Our findings show the importance of accounting for worker needs from the earliest stages of design. 
As Suchman noted, ``too often, assumptions are made as to how tasks are performed rather than unearthing the underlying work practices'' \cite{suchman_making_1995}. 
By learning from incidents, developers can create AI systems aligned with worker needs.
\clearpage
\section{Endmatter statements}
\label{sec:statements}





\subsection{Generative AI Usage Statement}
In this paper we made use of generative AI for three tasks, data analysis, code assistant, and editing, which we explain in detail next. 

\begin{enumerate}
    \item \textbf{Data analysis.} We used Open AI GPT-5.1 to assist us with data classification. This was used to identify workplace incidents, extract their job tasks, and analyse the misaligned AI traits that contributed to the incident. We explain all the steps in the Research Design section (\S\ref{sec:methods}).
    \item \textbf{Code assistant.} We used the free version of ChatGPT by OpenAI and Claude by Anthropic to assist with coding for creating the plots and the Overleaf tables. We followed similar practices than those from Stack Overflow, by looking for help when necessary, rather than generating all the content.
    \item \textbf{Editing.} Finally, we used again the free versions of ChatGPT and Claude to assist with text editing. This was minimally used in a way in which the full text was not rewritten, but rather we asked for recommendations on \textit{which} words to cut down, and which sentences might come across as repetitive.
\end{enumerate}




\subsection{Ethical Considerations Statement}
This research studied AI systems in the workplace and the misalignment between AI system traits and worker needs. All the data we used was drawn from publicly available sources, including the AI Incident Database \cite{mcgregor_preventing_2021} and prior user studies \cite{ranjit_are_we}, and no personally identifiable information was collected or analyzed. 
Our analysis emphasizes ethical AI design by highlighting how developers’ design choices can, even if not intentionally, contribute to workplace harms, with the aim of supporting safer and more worker-centered AI systems.

\bibliographystyle{ACM-Reference-Format}
\bibliography{workplace}

\appendix

\section{Appendix}


\subsection{LLM Rubric: Prompt to Classify Workplace-related Incidents}
\label{appendix:prompt_workplace}

\vspace{0.65cm}
\tikzstyle{background rectangle}=[thick, draw=black, rounded corners]
\begin{tikzpicture}[show background rectangle]
\node[align=justify, text width=45em, inner sep=1em]{
    \scriptsize 
    
    \noindent\textbf{System:} You are an experienced researcher studying how AI systems augment job tasks in the workplace. \\
    
    \noindent\textbf{User:} You will see an incident ID, TITLE, and DESCRIPTION. \\

Return True **only if** the incident describes a worker (contracted, self-employed, gig-worker), or an organisation (profit, non-profit, government, gig-platform) 
using AI **to augment, automate or manage a job task**. 
A task is a unit of work activity that produces a meaningful output.
The physical location does not matter: remote, field, factory, and office-based work all count as workplace contexts. 
\\

Else, return False.
\\

Exclude:\\
- Incidents where AI affects only end users, customers, or the public, without showing impact on workers performing job tasks.\\
- Social media, platform, or recommendation system errors unless they clearly affect workers performing their jobs.\\
- AI used for illegal, adversarial, or malicious purposes (e.g., fraud, blackmail, jailbreaking for illicit content).
\\

Output a JSON object with a key of incident ID and its value, and a key of workplace and its value of True or False, e.g.:
\\

\texttt{\{\{} \\
``incident\_id'': ``18'', \\
``workplace'': ``False'' \\
\texttt{\}\}}

-
\\

The ID of the incident: \texttt{\{incident\_id\}} \\
The title of the incident: \texttt{\{title\}} \\
The description of the incident: \texttt{\{description\}} \\

Whether the incident occurred in the workplace: \\        
};
\node[xshift=0.5ex, yshift=1ex, overlay, fill=black, text=white, draw=black, rounded corners, right=0.95cm, below=-0.3cm, inner xsep=0.55em, inner ysep=0.32em] at (current bounding box.north west) {
\textit{Prompt}
};
\end{tikzpicture}

\subsection{LLM Rubric: Prompt to Extract Job Tasks}\label{appendix:prompt_tasks}

\vspace{0.65cm}
\tikzstyle{background rectangle}=[thick, draw=black, rounded corners]
\begin{tikzpicture}[show background rectangle]
\node[align=justify, text width=45em, inner sep=1em]{
    \scriptsize 
    
    \noindent\textbf{System:} You are an experienced researcher studying how artificial intelligence tools augment human work tasks across different sectors. \\
    
    \noindent\textbf{User:}
You will receive a description of an AI incident and a list of job tasks (with IDs) in the industry sector of \texttt{\{\{sector\}\}}.
\\

First, determine whether the incident involves an AI system augmenting/automating workplace activities within a position in \texttt{\{\{sector\}\}} or within the sector of \texttt{\{\{sector\}\}}. 
If yes, proceed to rate each task below. If no, assign a rating of 0 to all tasks.
\\
                      
For each task in the \texttt{\{\{list\_of\_tasks\}\}}, imagine an AI system augmenting or automating this specific job task. Answer to which extent this incident description mentions such an AI system by rating it as below.
\\

Ratings: \\
\textbf{3} The incident mentions the AI system augmenting the task and its application in the task was the main source of the incident.\\
\textbf{2} The incident mentions the AI system augmenting the task and its application in the task was a partial source in the incident.\\
\textbf{1} The incident mentions the AI system augmenting the task but it is unclear whether the its application in task might have caused the incident.\\
\textbf{0} The incident does not mention the AI system augmenting the task.
\\               

Your output format should be:\\
A JSON object where each key is the task ID and the value is an object with the task description (as a string) and an object with "rating", e.g.
\\

\texttt{\{\{\\
  ``1''\: \{``task\_description''\: ``Customer data analysis'', ``rating''\: 0\}, \\
  ``2''\: \{``task\_description''\: ``Interpret data and regulations'', ``rating''\: 1\},\\
  ``3''\: \{``task\_description''\: ``Gather and analyse data", ``rating''\: 3\}\\
\}\}}

-
\\

Incident description: \texttt{\{\{incident\_description\}\}}

Incident sources: \texttt{\{\{incident\_sources\}\}}

List of job tasks with IDs, e.g., 1. Consumer data analysis:
\texttt{\{\{list\_of\_tasks\texttt\}\}}
\\

Question:
On a scale from 0 (incident does not mention AI system) to 3 (incident mentions and involves AI system augmenting the task), rate the extent to which this incident involves AI augmenting/automating each specific task in \texttt{\{\{sector\}\}}.    
};
\node[xshift=0.5ex, yshift=1ex, overlay, fill=black, text=white, draw=black, rounded corners, right=0.95cm, below=-0.3cm, inner xsep=0.55em, inner ysep=0.32em] at (current bounding box.north west) {
\textit{Prompt}
};
\end{tikzpicture}

\subsection{LLM Rubric: Prompt to Extract Misaligned Psychological Traits}\label{appendix:prompt_traits}

\vspace{0.65cm}
\tikzstyle{background rectangle}=[thick, draw=black, rounded corners]
\begin{tikzpicture}[show background rectangle]
\node[align=justify, text width=45em, inner sep=1em]{
    \scriptsize 
    
    \noindent\textbf{System:} You are an experienced researcher studying how AI systems augment job tasks in the workplace. \\
    
    \noindent\textbf{User:} Your task:

You are given a workplace AI incident.
The incident involves a worker using AI to augment a job task.

Your goal is to determine whether the incident involved TRAIT MISALIGNMENT:
i.e., the AI system behaved differently from what the worker needed.

For EACH label below, output ``Yes'' or ``No''.

-

TRAIT MISALIGNMENT LABELS

1. Complex vs. routine work \\
- complex\_sufficient: AI handled complex work, but worker needed routine work.\\
- complex\_necessary: Would the incident still occur if AI were NOT handling complex work?\\
- routine\_sufficient: AI handled routine work, but worker needed complex work.\\
- routine\_necessary: Would the incident still occur if AI were NOT handling routine work?\\

2. Emotions vs. apathetic\\
- emotion\_sufficient: AI focused on addressing human needs and emotions, but worker just needed data handling.\\
- emotion\_necessary: Would the incident still occur if AI were NOT focusing on addressing emotions?\\
- apathetic\_sufficient: AI just focused on data handling, but worker needed it to address human needs/emotions.\\
- apathetic\_necessary: Would the incident still occur if AI were NOT just focusing on data handling?\\

3. Fast vs. explainable\\
- fast\_sufficient: AI made fast, automatic decisions without explanation, but worker needed decisions that are easy to understand.\\
- fast\_necessary: Would the incident still occur if AI were NOT making fast, automatic decisions without explanation?\\
- explainable\_sufficient: AI was making decisions that are easy for people to understand, but worker needed fast, automatic decisions without explanations.\\
- explainable\_necessary: Would the incident still occur if AI were NOT making decisions that are easy for people to understand?\\

4. Open to challenge vs. definitive\\
- open\_sufficient: AI was open to challenge, but worker needed a final decision.\\
- open\_necessary: Would the incident still occur if AI were NOT open to challenge?\\
- definitive\_sufficient: AI treated decisions as final, but worker needed openness to challenge.\\
- definitive\_necessary: Would the incident still occur if AI were NOT treating decisions as final?\\

5. Personalised vs. generalised\\
- personalised\_sufficient: AI adjusted based on the individual it was helping, but the worker needed it to treat everyone similarly.\\
- personalised\_necessary: Would the incident still occur if AI were NOT adjusting based on the individual it was helping?\\
- generalised\_sufficient: AI treated everyone similarly, but the worker needed it to adjust based on the individual.\\
- generalised\_necessary: Would the incident still occur if AI were NOT treating everyone too similarly?\\

6. Warm vs. business-like\\
- warm\_sufficient: AI showed warmth and care, but the worker needed it to remain neutral and business-like.\\
- warm\_necessary: Would the incident still occur if AI were NOT showing warmth and care?\\
- businesslike\_sufficient: AI remained neutral and business-like, but the worker needed it to show warmth and care.\\
- businesslike\_necessary: Would the incident still occur if AI were NOT remaining neutral and business-like?\\

7. Polite vs. straightforward\\
- polite\_sufficient: AI was polite even if that meant not being fully honest, but the worker needed it to be sincere and straightforward.\\
- polite\_necessary: Would the incident still occur if AI were NOT being polite even if that meant not being fully honest?\\
- straightforward\_sufficient: AI was sincere and straightforward, but the worker needed it to be polite even if that meant not being fully honest.\\
- straightforward\_necessary: Would the incident still occur if AI were NOT being sincere and straightforward?\\

8. Tolerant/open-minded vs. strict\\
- tolerant\_sufficient: AI was tolerant and open-minded, but the worker needed it to be strict and follow rules exactly.\\
- tolerant\_necessary: Would the incident still occur if AI were NOT being tolerant and open-minded?\\
- strict\_sufficient: AI strictly followed rules, but the worker needed it to be tolerant and open-minded.\\
- strict\_necessary: Would the incident still occur if AI were NOT strictly following rules?\\

};
\node[xshift=0.5ex, yshift=1ex, overlay, fill=black, text=white, draw=black, rounded corners, right=0.95cm, below=-0.3cm, inner xsep=0.55em, inner ysep=0.32em] at (current bounding box.north west) {
\textit{Part 1/2}
};
\end{tikzpicture}

\vspace{0.65cm}
\tikzstyle{background rectangle}=[thick, draw=black, rounded corners]
\begin{tikzpicture}[show background rectangle]
\node[align=justify, text width=45em, inner sep=1em]{
    \scriptsize 

9. Precise vs. basic\\
- precise\_sufficient: AI was highly skilled and precise, but the worker needed it to be fast and simple even if less perfect.\\
- precise\_necessary: Would the incident still occur if AI were NOT being highly skilled and precise?\\
- basic\_sufficient: AI was fast and simple even if less perfect, but the worker needed it to be highly skilled and precise.\\
- basic\_necessary: Would the incident still occur if AI were NOT being fast and simple even if less perfect?\\

10. Flexible vs. determined\\
- flexible\_sufficient: AI was flexible and willing to change course, but the worker needed it to be determined and persistent.\\
- flexible\_necessary: Would the incident still occur if AI were NOT being flexible and willing to change course?\\
- determined\_sufficient: AI was determined and persistent, but the worker needed it to be flexible and willing to change course.\\
- determined\_necessary: Would the incident still occur if AI were NOT being determined and persistent?\\

11. Insightful/comprehensive vs. simple\\
- insightful\_sufficient: AI showed comprehensiveness, deep understanding, and insight, but the worker needed it to keep things simple and straightforward.\\
- insightful\_necessary: Would the incident still occur if AI were NOT showing comprehensiveness, deep understanding, and insight?\\
- simple\_sufficient: AI kept things simple and straightforward, but the worker needed it to show comprehensiveness, deep understanding, and insight.\\
- simple\_necessary: Would the incident still occur if AI were NOT keeping things simple and straightforward?\\

12. Practical vs. imaginative\\
- practical\_sufficient: AI stayed practical and followed familiar approaches, but the worker needed it to be imaginative and bring new ideas.\\
- practical\_necessary: Would the incident still occur if AI were NOT staying practical and following familiar approaches?\\
- imaginative\_sufficient: AI was imaginative and brought new ideas, but the worker needed it to stay practical and follow familiar approaches.\\
- imaginative\_necessary: Would the incident still occur if AI were NOT being imaginative and bringing new ideas?\\

---

OUTPUT FORMAT:

\texttt{\{\{} \\
  ``complex\_sufficient'': ``Yes\textbar No'', \\
  ``complex\_necessary'': ``Yes\textbar No'', \\
  ``routine\_sufficient'': ``Yes\textbar No'', \\
  ... \\
  ``practical\_necessary'': ``Yes\textbar No'', \\
  ``imaginative\_sufficient'': ``Yes\textbar No'', \\
  ``imaginative\_necessary'': ``Yes\textbar No'' \\
\texttt{\}\}}

---

INCIDENT DETAILS \\

INCIDENT\_ID: \texttt{\{incident\_id\}} \\
TASK: \texttt{\{task\_description\}} \\
TITLE: \texttt{\{title\}} \\
DESCRIPTION: \texttt{\{description\}} \\
SOURCES: \texttt{\{sources\}} \\

***Check your planned output before outputting it: if it contains any explanations besides the JSON string, omit the explanations. Make sure to output ONLY a correctly formatted JSON string and nothing else. Do not miss any of the traits.***
};
\node[xshift=0.5ex, yshift=1ex, overlay, fill=black, text=white, draw=black, rounded corners, right=0.95cm, below=-0.3cm, inner xsep=0.55em, inner ysep=0.32em] at (current bounding box.north west) {
\textit{Part 2/2}
};
\end{tikzpicture}

\newpage
\section{Survey Details}\label{appendix:developers}
\begin{table}[h!]
\centering
\caption{Characteristics of participants recruited from Prolific, recruited from \cite{ranjit_are_we}. The workers were highly familiar with the tasks in their corresponding sector. Developers work in software, data, IT, and ML/AI roles who actively engage with modern AI tools and contribute to AI enabled workflows across diverse organizational sectors.}
\label{tab:demographics}
\small
\begin{tabular}{lcc}
\hline
\textbf{Demographics} & \textbf{Workers (N=202)} & \textbf{Developers (N=197)} \\
\hline
Mean age (SD) & 42.63 (13.05) & 36.68 (10.29) \\
\hline
\textbf{Gender} & & \\
Male & 34.03\% & 64.46\% \\
Female & 61.94\% & 28.42\% \\
Non-binary / Other & 0\% & 0\% \\
Consent Revoked & 4.03\% & 7.11\% \\
\hline
\textbf{Employment status} & & \\
Full-time & 46.94\% & 69.54\% \\
Part-time & 16.29\% & 13.71\% \\
Unemployed / Other & 4.03\% & 1.52\% \\
Consent Revoked / No data available & 15.22\% & 23.80\% \\
\hline
\end{tabular}
\end{table}

\begin{table}[h]
\centering
\caption{Survey items on preference of AI traits}
\label{tab:ai_traits_questions}
\small
\begin{tabular}{p{0.05\linewidth} p{0.56\linewidth}}
\toprule
\textbf{\#} & \textbf{Item} \\
\midrule
1 & Handle more complex work rather than routine work. \\
2 & Focus more on addressing human needs and emotions rather than just data handling. \\
3 & Make fast, automatic decisions without explanation rather than decisions that are easy for people to understand. \\
4 & Be open to challenge or treat the decision as final. \\
5 & Adjust based on the individual it’s helping rather than treat everyone the same. \\
6 & Show warmth and care rather than remain neutral and business-like. \\
7 & Be polite even if that means not being fully honest, rather than being sincere and straightforward. \\
8 & Be strict and follow the rules exactly rather than be tolerant and open-minded. \\
9 & Be fast and simple even if less perfect, rather than highly skilled and precise. \\
10 & Be determined and persistent rather than flexible and willing to change course. \\
11 & Show comprehensiveness, deep understanding and insight rather than keep things simple and straightforward. \\
12 & Be imaginative and bring new ideas rather than stay practical and follow familiar approaches. \\
\bottomrule
\end{tabular}
\end{table}

\begin{table}[h]
\centering
\caption{Definitions of sectors, adapted from \cite{ranjit_are_we}.}
\label{tab:sector_definitions}
\footnotesize
\begin{tabular}{p{0.28\textwidth} p{0.68\textwidth}}
\hline
\textbf{Sector} & \textbf{Definition} \\
\hline
\textbf{Operations} & Activities concerned with planning, coordinating, and executing an organization’s core processes for producing goods or delivering services, including workflow management, logistics, pricing, and operational administration. \\

\textbf{Human Resources} & Functions related to managing an organization’s workforce, including recruitment, employment recordkeeping, workforce analytics, and administration of employee lifecycle events in accordance with organizational and regulatory requirements. \\

\textbf{Finance or Accounting} & Activities focused on designing, implementing, and maintaining financial and accounting systems to track, manage, and report economic transactions, ensuring financial accuracy, regulatory compliance, and decision support. \\

\textbf{Engineering (e.g., Software)} & Technical activities involving the design, development, modification, supervision, and maintenance of engineered systems, particularly software systems, to ensure functionality, performance, and alignment with technical specifications. \\

\textbf{Data Analysis} & The systematic cleaning, processing, modeling, visualization, and interpretation of data to identify patterns, trends, and relationships that support evidence-based decision-making. \\

\textbf{Research} & Systematic investigation involving data collection, literature review, survey methods, and statistical analysis to generate, validate, and extend knowledge for scientific, policy, or organizational purposes. \\

\textbf{Healthcare Professional} & Work focused on the assessment, treatment, monitoring, and coordination of patient care, including clinical decision-making, documentation, patient advocacy, and supervision to support physical, mental, and social well-being. \\

\textbf{Legal} & Activities involving the interpretation, application, and analysis of laws and regulations, including legal research, document preparation, advisory services, case strategy development, and advocacy. \\

\textbf{Education Professional} & Activities related to designing, delivering, and evaluating instructional practices, including classroom management, student assessment, curriculum implementation, and use of educational technologies to support learning and development. \\

\textbf{Design or Creative} & Work centered on the conceptualization and production of visual, written, or artistic content, translating ideas, narratives, and aesthetic principles into tangible outputs using artistic and digital tools. \\

\textbf{Public Relations / Communications} & Activities focused on managing information flows between organizations and their audiences through strategic messaging, media relations, content creation, and reputation management. \\

\textbf{Customer Experience / Support} & Direct customer-facing activities aimed at providing information, resolving service or billing issues, and maintaining customer satisfaction through communication and problem resolution. \\

\textbf{Account Management} & Activities involving the administration and support of customer financial relationships, including processing applications, explaining financial products, and managing ongoing accounts to ensure accuracy, compliance, and service continuity. \\
\hline
\end{tabular}
\end{table}

\newpage
\section{Chronological Analysis}\label{appendix:chronological}
\begin{figure}[H]
    \centering
    \includegraphics[width=1\columnwidth]{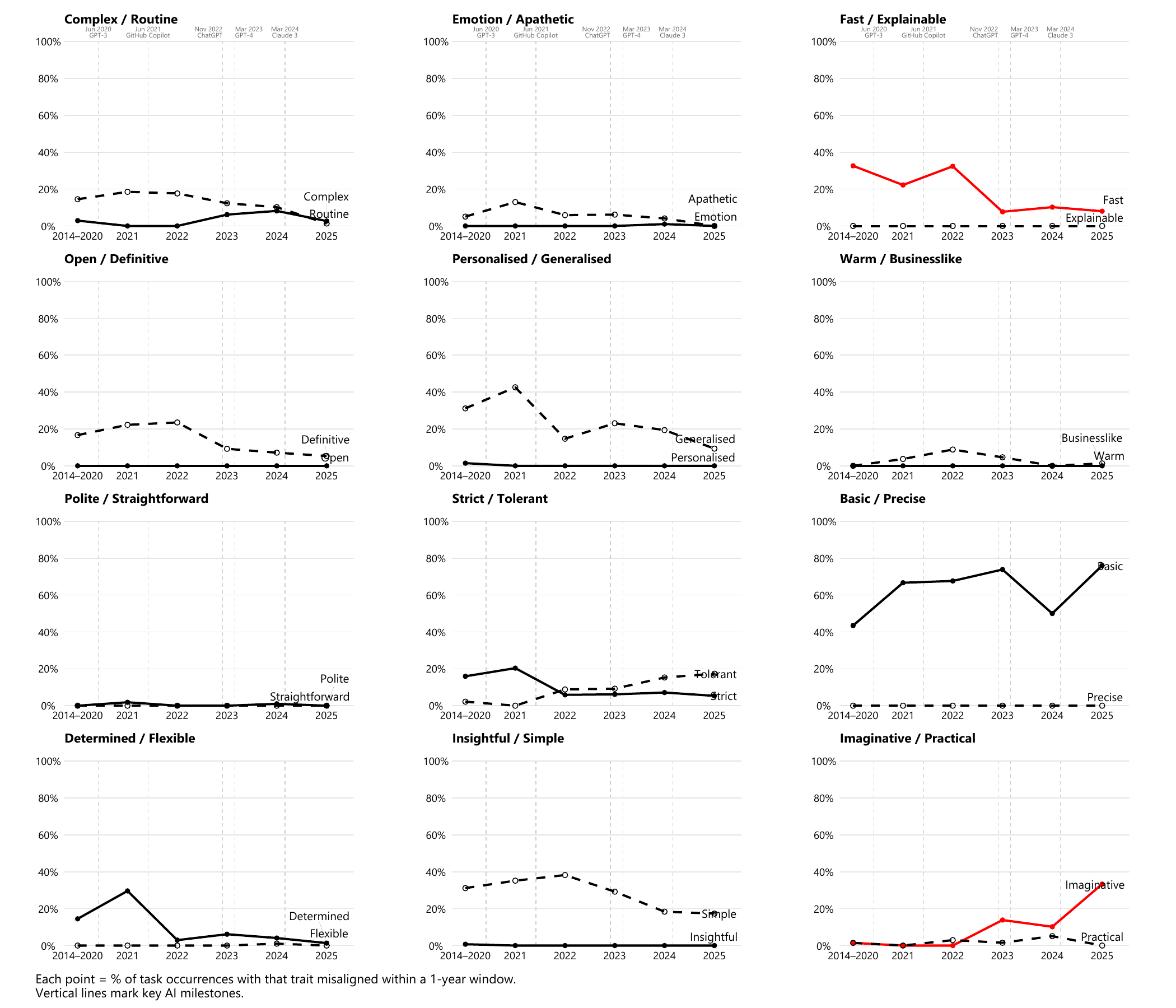}
    \caption{
    \textbf{
    Percentage of task occurrences involving incidents of misaligned AI for each pair of traits over time.} 
Each data point represents the percentage of task occurrences classified as specified in the label within a year. We merged 2014-2020 as there were significantly fewer data points. For each pair of traits, the first trait listed (e.g., complex) is represented by the solid line, and the second trait listed (e.g., routine) is represented by the dashed line. Vertical lines mark milestones in AI research and deployment for context \cite{standford2025ai}. Interestingly, the frequency of misalignments involving specific traits increased or decreased over time: fast AI used to cause incidents, yet, since 2022, imaginative AI started to do so.}
    \label{appendix:traits_time}
\end{figure}

\newpage
\begin{table*}[t]
\centering
\caption{Distribution of incidents that happened because of trait misalignment or other reason.}
\small
\label{tab:misaligned_incidents}
\begin{tabular}{ccc}
\toprule
\# Type & \# incidents & \% incidents\\
\midrule
Misaligned  & 179 & 83.4\\
Aligned  & 35 & 16.6\\
\bottomrule
\end{tabular}
\end{table*}

\begin{table*}[t]
\centering
\caption{Distribution of the number of trait misalignments per incident.}
\small
\label{tab:causal_counts}
\begin{tabular}{ccc}
\toprule
\# trait misalignments & \# incidents & \% incidents\\
\midrule
0  & 37 & 17.3\\
1  & 32 & 15.0\\
2  & 33 & 15.4\\
3  & 26 & 12.1\\
4  & 24 & 11.2\\
5  & 22 & 10.3\\
6  & 19 & 8.88\\
7  & 12 & 5.61\\
8  & 4  & 1.87\\
9  & 4  & 1.87\\
10 & 1  & 0.47\\
\bottomrule
\end{tabular}
\end{table*}

\end{document}